\documentclass[prc,aps,preprint,showpacs]{revtex4-1}
\usepackage{ae,aecompl}
\usepackage[T1]{fontenc}
\usepackage[latin9]{inputenc}
\usepackage{amsmath}
\usepackage{amssymb}
\usepackage{graphicx}
\usepackage{multirow}
\usepackage{mathtools}
\usepackage{subfigure}

\makeatletter

\pdfpageheight\paperheight
\pdfpagewidth\paperwidth

\newcommand{\noy}[3]{$\prescript{#2}{#3}{\mathrm{#1}}$}
\newcommand{\noym}[3]{\prescript{#2}{#3}{\mathrm{#1}}}

\newcommand{\E}{\mathrm{E}}

\newcommand{\Z}{\mathrm{Z}}
\newcommand{\A}{\mathrm{A}}

\newcommand{\R}{\mathrm{R}}
\newcommand{\T}{\mathrm{T}}
\newcommand{\p}{\mathrm{p}}
\newcommand{\e}{\mathrm{e}}

\begin{document}
 
\title{A model for particle production in nuclear reactions at intermediate energies:\\
 application to C-C collisions at 95~MeV/nucleon}

\author{J. Dudouet}
\author{D. Durand}
\affiliation{LPC Caen, ENSICAEN, Universit\'e de Caen, CNRS/IN2P3, Caen, France}

%\date{\today}
\begin{abstract}
A model describing nuclear collisions at intermediate energies is presented and the results are compared with recently measured double differential cross sections in C-C reactions at 95~MeV/nucleon.  Results show the key role played by geometrical effects and the memory of the entrance channel, in particular the momentum distributions of the two incoming nuclei. Special attention is paid to the description of processes occurring at mid-rapidity. To this end, a random particle production mechanism by means of a coalescence process in velocity space is considered in the overlap region of the two interacting nuclei. 
\end{abstract}                           


\maketitle

\section{Introduction}

In a previous paper~\cite{Dudouet14a}, the available nuclear collision models implemented in the GEANT4 toolkit~\cite{GEANT4} have been "benchmarked" with help of a comparison with experimental data obtained in C-C reactions at 95~MeV/nucleon~\cite{Dudouet13b,Dudouet14b}. Results of this work showed discrepancies between the models and the data. In particular, it was pointed out some difficulties to reproduce correctly the so-called mid-rapidity region, i.e the kinematical region in-between the projectile and the target velocities. The models implemented in GEANT4 are all dynamical models often coupled to an evaporation model that treat secondary decays. The discrepancies at mid-rapidity are linked to the difficulty of reproducing the production and the kinematics of light cluster produced by the mixing of nucleons originated from both projectile and target. In the present work, we present a model called SPIIPIE (Simulations of Light Ions Induced Processes at Intermediate Energies) with some hypotheses that could help improving this point. In particular, we consider a scenario in which clusters are produced very rapidly well before the system reaches equilibrium. This is achieved by considering an almost ``instantaneous'' aggregation based on the initial conditions of the reaction.

\addtocontents{toc}{\vspace{0.4cm}}
\section{Model assumptions}

A main feature of the model is based on a strict geometrical assumption similar to the so-called participant-spectator model which is widely used at higher energies. This means that for a given impact parameter, the nucleons are shared among three different species, the quasi-projectile (-target) made of nucleons belonging to the projectile (target) and not belonging to the overlap geometrical region between the two partners and the participant zone made of nucleons belonging to the overlap region.

We also consider a two-step semi-microscopic model related to two different time scales of the reaction. In a first short step, the so-called entrance channel, particle and excited fragment production occur. In a second step, the exit channel, on a larger time scale (typically of several tens of fm/c and at times much larger than the reaction time), excited species decay by particle emission and this is considered using the usual statistical decay theory. By semi-microscopic, it is meant that the degrees of freedom of the model are considered both at microscopic and macroscopic levels. At the microscopic level, the internal momentum and spatial distributions of the nucleons of the two incoming nuclei are considered. At the macroscopic level, the collision is simulated by means of geometrical assumptions and macroscopic quantities such as excitation energies are estimated using ``by hand'' prescriptions. At variance with fully dynamical models such as the intra-nuclear cascade or quantum molecular dynamics approaches which consider the time evolution of the ensemble of nucleons, a major aspect of the present model is based on what we could call a sudden and frozen approximation. It is indeed assumed that particle and fragment production occurs on such a very short time scale so that the momentum distributions of the incoming nuclei have no time to fully relax and are only affected initially by a given amount of hard in medium nucleon-nucleon collisions. As such, the kinematical characteristics of the particle and the fragments at their time of production is entirely determined by the almost unperturbed and frozen initial nucleon distribution and thus keep a strong memory of the entrance channel of the reaction.

\subsection{Entrance channel modelisation}

\subsubsection{Initial conditions}

The initialization procedure consists in preparing the two incoming nuclei before treating the collision in itself. In particular, the internal momentum distribution is built based on well known shell model distributions:
\begin{equation}
\label{equ:OH}
 \frac{dN}{d\p} = \left[ 1+ \frac{(\A-4)\times b_0 \times \p^2}{6} \right] \times \p^2\e^{-b_0\p^2},
\end{equation}
where  A is the mass and $b_0=68.5\times10^{-6}$~(MeV/c)$^2$~\cite{Golubeva93,Daskalov90}.

Similar distributions are used for both proton and neutron. A center-of-mass correction is applied to ensure that the initial nuclei are at rest. Then, the projectile distribution is boosted in the laboratory frame.   

Each impact parameter, $b$, is sampled between $b=0$ and $b_{max}$, $b_{max}$ being the sum of the radius, $\R$, of two partners of the reaction~\cite{Wollersheim05}: 
\begin{equation}
\label{equ:Rayon}
 \R = 1.28\A^{1/3} - 0.76 + 0.8\A^{-1.3}~~(\text{fm}),
\end{equation}
where $\A$ is the mass number of the nucleus. The overlap function of the projectile and the target is calculated by Monte Carlo based on well known density distributions. From the overlap function, it is thus possible to determine the size of the quasi-projectile $A_{QP}$ and of the quasi-target $A_{QT}$. The remaining nucleons constituting the participant zone (PZ). Let us call $A_\text{Proj}^{PZ}$ and $A_\text{Targ}^{PZ}$ the number of nucleons in PZ respectively from the projectile and from the target. We have $A^{PZ} = A_\text{Proj}^{PZ} + A_\text{Targ}^{PZ}$. Before considering the process by which these $A^{PZ}$ nucleons aggregate to form clusters and free particles, one has first to take into account the fact that their momentum distribution can be modified by in-medium nucleon-nucleon collisions.

%Figure~\ref{fig:OverlapProp} shows the results 
%for several systems (\noy{C}{12}{}+\noy{H}{1}{}, \noy{C}{12}{}+\noy{C}{12}{}, et \noy{C}{12}{}+\noy{Ti}{48}{}.%

%\begin{figure}[]
%\centering \includegraphics[width=0.5\linewidth]{img/OverlapProp/OverlapProp.pdf}
%\caption{Overlap functions (see text) for \noy{C}{12}{}+\noy{H}{1} (blue) {}, \noy{C}{12}{}+\noy{C}{12} (green) {}, et \noy{C}{12}{}+\noy{Ti}{48} (red){}}
%\label{fig:OverlapProp}
%\end{figure}%

\subsubsection{In medium nucleon nucleon collisions}

\begin{figure*}[!ht]
\centering \includegraphics[width=0.8\linewidth]{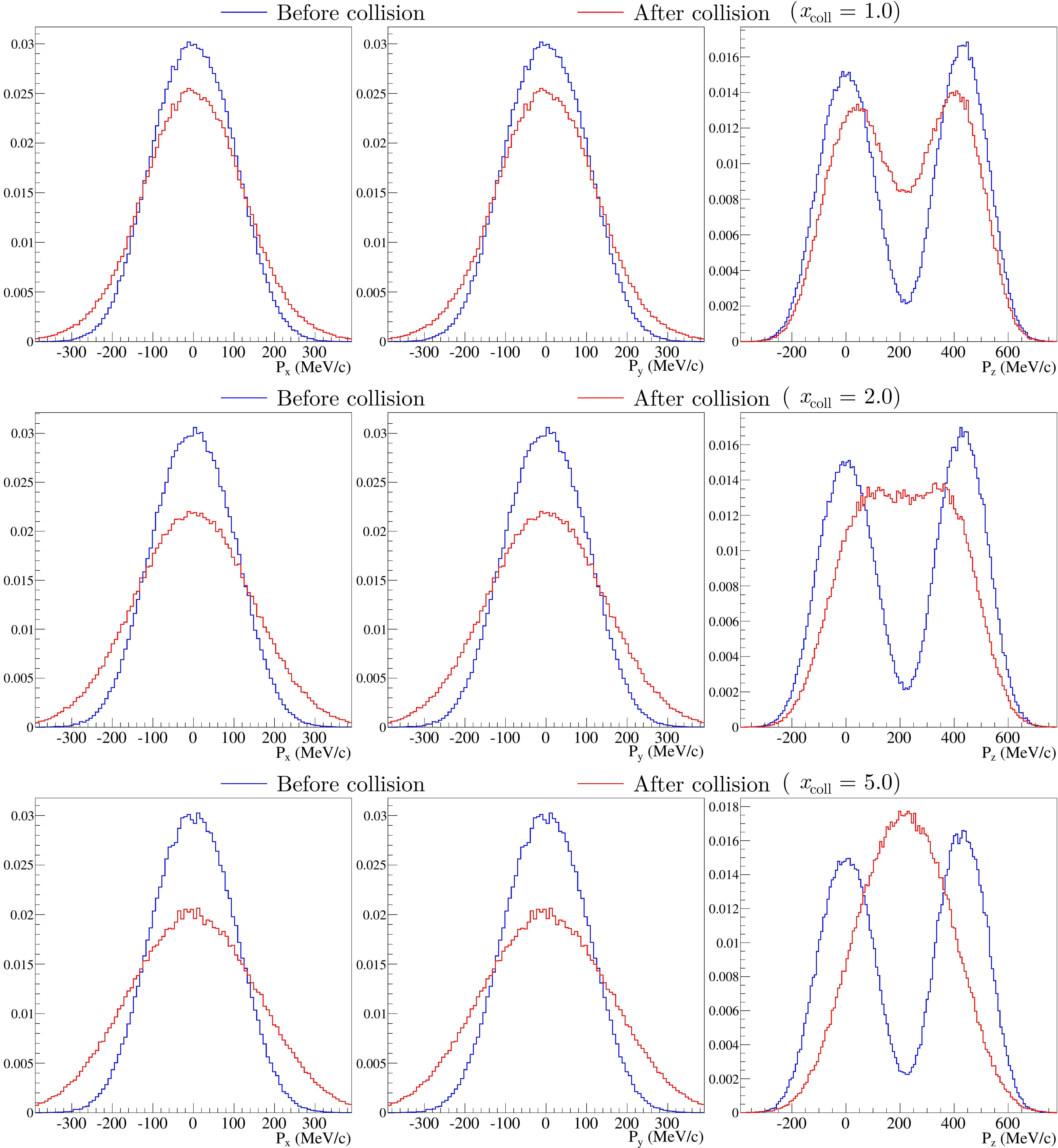}
\caption{(Color online) Normalized momentum distributions of participant nucleons before (blue) and after (red) nucleon-nucleon collisions for different values of $x_\text{coll}$ . Left: x-axis
middle: y-axis and right:  z-axis corresponding to the beam axis.}
\label{fig:Coll}
\end{figure*}

\begin{figure}[!ht]
\centering \includegraphics[width=0.6\linewidth]{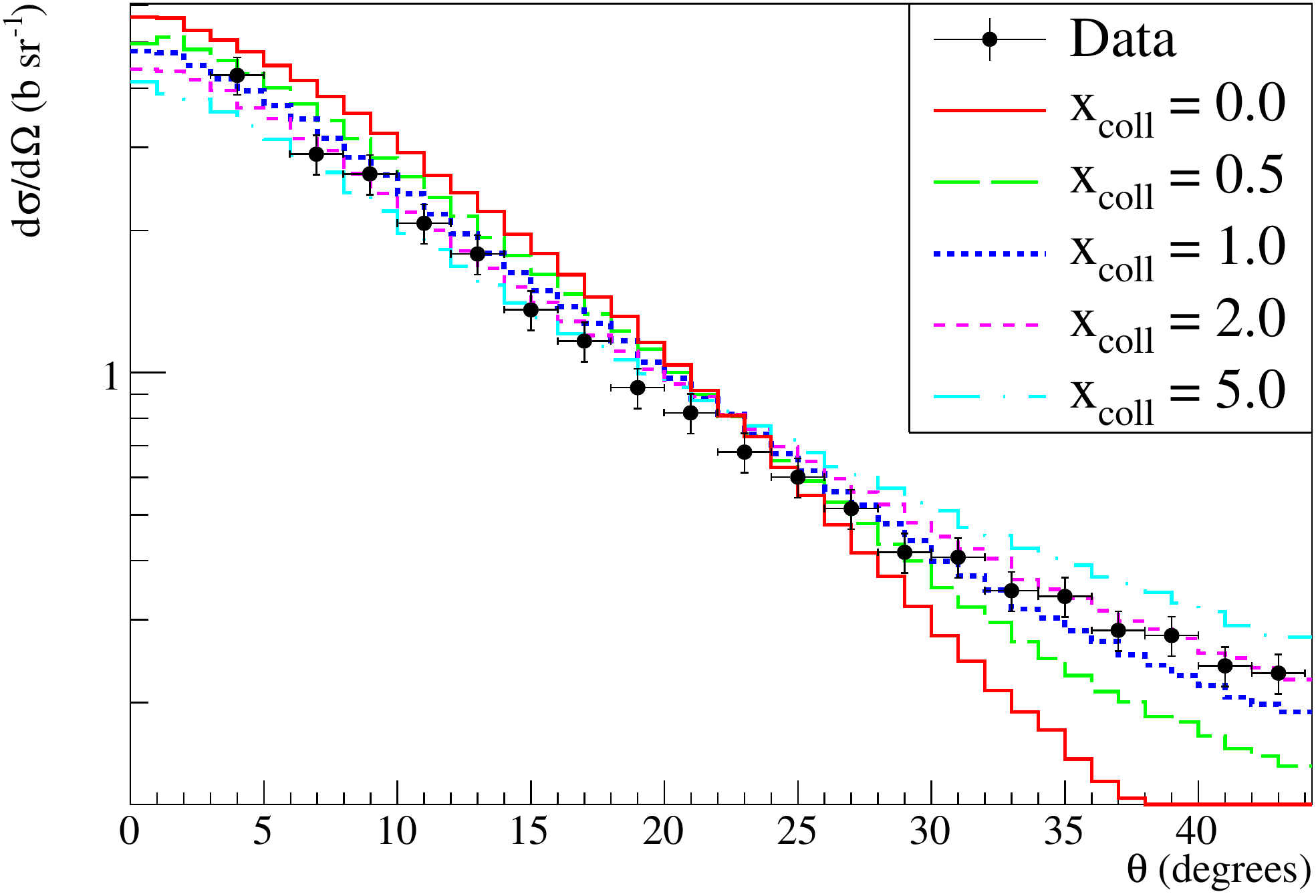}
\caption{(Color online) Angular distributions of the protons for the reaction \noy{C}{12}{}+\noy{C}{12}{} at 95~MeV/nucleon for different values of $x_\text{coll}$.}
\label{fig:Coll_AngDist}
\end{figure}

Indeed, at intermediate energies, nucleons belonging to the participant zone (PZ) experience hard in-medium nucleon nucleon collisions. Let us call $x_\text{coll}$ the average number of collisions per participant from the incoming projectile, we thus have for the total number of collisions:
\begin{equation}
 N_\text{coll} = x_\text{coll} \times min(A_\text{Proj}^{PZ},A_\text{Targ}^{PZ})
\end{equation}
For a symmetric system considered here, $A_\text{Proj}^{PZ}=A_\text{Targ}^{PZ}$. The $N_\text{coll}$ collisions are treated by considering at random a nucleon from the projectile and a nucleon from the target belonging to the participant zone. The elastic scattering is then performed using fitted 
free nucleon-nucleon angular distribution. For p+p and n+n collisions, the distribution is isotropic while for n+p, the differential cross-section is borrowed from~\cite{Li93}:
\begin{align*}
 \frac{d\sigma}{d\Omega}(E_{lab},\theta) &= \frac{17.42}{1+0.05\times \left( \E_{lab}^{0.7} - 15.5 \right)} \times \exp{ \left[\alpha \left(cos^2\theta + sin^2\frac{\theta}{7} - 1.0 \right)\right]},\\
\notag \text{with }\alpha &= 0.125 \left( E_{lab}^{0.54} -4.625\right)\text{for }E_{lab}\leq 100~MeV,\\
\notag \text{and }\alpha &= 0.065 \left(36.65  -  E_{lab}^{0.58} \right)\text{for }E_{lab} > 100~MeV.
\end{align*}
Note that the value of  $\alpha$ quoted in the original paper is wrong and has been corrected here in agreement with the authors. Figure~\ref{fig:Coll} shows the momentum distributions of the nucleons in the overlap zone (here in the case $b=0$) before (blue) and after (red) collisions for various values of $x_\text{coll}=1.0$, 2.0 and 5.0. 
The longitudinal distributions along the axis $\overrightarrow{u_z}$ displayed in the lab frame show a slight asymmetry between the projectile and the target. This is a relativistic effect
due to the contraction of the momentum distribution along the beam axis.  The transverse distributions, $\p_x$ et $\p_y$ are enlarged and the longitudinal distributions $\p_z$ are damped and may overlap although for the lower values of $x_\text{coll}$ the two distributions keep a memory of the entrance channel meaning that full thermalization has not occurred. However, for $x_\text{coll}=5.0$, the two distributions totally overlap. 
It turns out that the number of collisions has little influence on the kinematics of fragments. The main effect concerns free nucleons. The value of $x_\text{coll}$ has been fixed by comparison with the measured proton angular distribution as displayed in Fig.~\ref{fig:Coll_AngDist}. The best agreement between the model and the experimental data is obtained for values of $x_\text{coll}$ between 1 and 2.

Although $x_\text{coll}$ is here considered as a free parameter, it is possible to have a rough estimation of its value on the basis of the free nucleon-nucleon cross section, $\sigma_{NN}^{free}$, using the following parametrization~\cite{kikuchi68}:
\begin{equation}
 \sigma_{nn} = \sigma_{pp} = \left( \frac{10.63}{\beta^2} - \frac{29.92}{\beta} + 42.9 \right)~\text{mb},~~~~\sigma_{np} = \sigma_{pn} = \left( \frac{34.10}{\beta^2} - \frac{82.2}{\beta} + 82.2 \right)~\text{mb},
\end{equation}
where $\beta$ is the reduced velocity of the nucleon. 

For E$_{\text{beam}}$=95~MeV/nucleon, $\sigma_{NN}^{free} = (\sigma_{nn}+\sigma_{np})/2 = (32.04+80.25)/2 = 56.15~\text{mb}$. The free cross section is corrected by the final state interaction also known as the Pauli blocking factor $\alpha_{Pauli}$ which can be parametrized as~\cite{kikuchi68}:
\begin{equation}
 \alpha_{Pauli} = 
  \left\{
 \begin{array}{l l}
  1-\frac{7}{5}\xi & ,\text{for } \xi\leq\frac{1}{2}\\
  1-\frac{7}{5}\xi+\frac{2}{5}\xi\left(2-\frac{1}{\xi}\right)^{5/2} & ,\text{for } \xi > \frac{1}{2}\\
 \end{array}
 \right.\
\end{equation}
where $\xi = E_F/E$, $E_F$=38~MeV is the Fermi energy and $E$ is the average kinetic energy of the incident nucleon (95~MeV). This gives $\alpha_{Pauli} = 0.44$ and $\sigma_{NN}^{medium} = \alpha_{Pauli} \times \sigma_{NN}^{free} = 24.8$~mb. 

For $b=0$, the number of collisions per incident nucleon writes:
\begin{equation} 
x_\text{coll} = \frac{1}{A}\int_0^{\R_\text{max}} 2\pi\text{rdr}\left( \sum_{i=1}^{A} i\e^{-N_\text{r}} \frac{N_\text{r}^i}{i!}\right) \times 2\rho_0\sqrt{\R_\text{max}^2-\text{r}^2}
\end{equation}
For $A=12$ (\noy{C}{12}{}+\noy{C}{12}{}), $\R_\text{max} = 2.52$~fm (cf.~equ.~\ref{equ:Rayon}), $\rho_0=0.168$~fm$^{-3}$ et $\sigma_{NN}^{free} = 24.8$~mb, we obtain $x_\text{coll} = 1.48$. Using $\sigma_{NN}^{free} = 28.5$~mb as in~\cite{Zeng96}, $x_\text{coll} = 1.70$. These values are compatible with the value obtained previously from data comparisons. In the following, the full comparison with the experimental is performed with $x_\text{coll} = 1.5$.

\subsubsection{Random coalescence between participants}
\label{sec:Coalescence}

Particle production in the overlap zone is now discussed. Many different algorithms have already been proposed in the literature \cite{Mashnik14,Toneev83,Schulz83,Mancusi2014,Kerby15} in the framework for instance of the Intra-Nuclear Cascade or  MCNP. Here, we consider a coalescence process in momentum space. The main idea is to use a stability criterion. This means that we only consider the production of clusters for which the internal relative kinetic momentum of the nucleon with respect to the fragment does not exceed a given value $p_{cut}$ which is a free parameter of the model. In other words, one starts by choosing randomly a nucleon among the $PZ$'s. Then, a second one is chosen and the relative momentum $p_{rel}$  is calculated. If  $p_{rel}$ is larger than $p_{cut}$, the aggregation does not occur and the nucleon is the seed of a new fragment. Otherwise, a deuteron is produced (in the procedure, we only consider ``realistic'' clusters thus excluding di-neutrons, di-protons and larger unstable clusters. The process is iterated by  assigning randomly each new nucleon in the process either to an already existing cluster (if possible) or to a new fragment until the end of the procedure. At the end of the process, all nucleons in the participant zone have been attributed to clusters or are considered as free nucleons. Note that we do not consider the aggregation process in real space but only in momentum space. The reason is that nucleons are delocalized and the geometrical size of the overlap region is of the order of the extension of the wavelength of the nucleons. In the following, the results of the model are shown with an optimized value value of $p_{cut} = 225~$MeV/c. It turns out that this value is close to the Fermi momentum ($\sim$250 MeV/c).

\subsubsection{Fragment excitation energy}

At the end of the coalescence process, the system is left in a state corresponding to a quasi-projectile, a quasi-target and several fragments or free nucleons in the mid-rapidity region. Such species are produced in excited states and the excitation energy has to be determined before considering the second step of the model, namely the deexcitation process. For such a light system as C-C, discrete known excited states are considered. Such states are assumed to be populated thermally and, as such a temperature has to be defined. As far as that the spectators are concerned, the prescription described in~\cite{Mallik11} is used. The temperature is a function of the impact parameter and writes: 
\begin{equation}
 T(b) =  T_{0} - T_{bmax}\left[\frac{A_{QP}(b)}{A_{Proj}}\right]~MeV.
\end{equation}
where $T_{0}=7.5~\text{MeV}$ is the temperature for b=0 and $T_{bmax}=4.5~\text{MeV}$ for $b_{max}$. Using an argument of continuity, $T_{0}$ is accordingly the temperature for the fragments in the overlap zone.
 
Writing the volume of a nucleus as:
\begin{equation}
 \notag V = \frac{4}{3}\pi r_0^3 A,
\end{equation}
one gets:
\begin{align}
\notag  T(b) &=  T_{0} - T_{bmax}\left[\frac{V_{QP}(b)}{V_{Proj}}\right]~MeV,\\
\label{equ:Temp}
 T(b) &=  T_{0} - T_{bmax}\times (1-f_{Proj}(b))~MeV,
\end{align}
where $f_{Proj}(b)$ is the impact parameter dependent overlap function as discussed previously.

Therefore, for each $b$, the excitation energy of the species is sampled by Monte Carlo using a thermal assumption:
\begin{equation}
 w_i = (2J^\pi_i+1)\e^{-E_i/T(b)},
\end{equation}
where the degeneracy and the energy of each discrete level is found in \url{http://www.nndc.bnl.gov/chart/}. Note that we thus only consider known discrete states. This is  probably at the origin of some discrepancies between the experimental data and the results of the calculation to be discussed later. It would be interesting to extend the level density distribution to higher energies using a continuum approximation as for example in~\cite{Baiocco12}, but, for the present work, such extension has not been done. 

At the end of the coalescence procedure, the total energy of the configuration is calculated by taking into account the mass defects, the excitation energies and the kinetic energies of the whole species. The momentum of each fragment is obtained by adding the momentum of each nucleon belonging to the fragment. Note that considering such a light system as C+C, Coulombic final state interaction between the species is not taken in to account.  

\subsubsection{Algorithm for the conservation of energy}

The procedure described above does not conserve the total energy of the system. A simple algorithm is thus used to respect energy conservation. An exchange process between the produced species is thus applied. Two protons or two neutrons are randomly chosen and exchanged among the fragments. The total energy of the configuration is recalculated
and the following quantity is minimized:
\begin{equation}
 X_{min} = \left|\frac{\noym{\Delta}{\A_{Targ}}{\Z_{Targ}} + \noym{\Delta}{\A_{Proj}}{\Z_{Proj}} + \T_{Proj}}{\sum_i\left(\noym{\Delta}{\A_{i}}{\Z_{i}} + \T_{i} + \E^*_i\right)}-1\right|
\end{equation}
where \noy{\Delta}{\A_{i}}{\Z_{i}} is the mass excess of the nucleus \noy{X}{\A_{i}}{\Z_{i}}, $\T_{i}$ the kinetic energy and $\E^*_i$ the excitation energy.

The process is iterated until $X_{min}$ is less than one percent.   

\subsection{Exit or decay channel modeling}

In the following, the decay model used to de-excite the species is briefly described. It was shown in our last paper~\cite{Dudouet14a} that the Fermi Break-Up model is by far the most suitable model for such a light system as C+C. Therefore, for each excited species produced in the entrance channel with excitation energy  $E^*$, all possible decay channels with species labeled $i$ that are energetically possible are considered. This procedure includes all possible discrete values for the excited states of each fragment $i$ . Let $n$ be the multiplicity of fragments in the considered decay channel. The energy balance writes for each partition:
\begin{equation}
E_{avail} = Q + E^* - \sum^n_i E^*_i
\end{equation}
 where  $E^*_i$ is one of the possible excited state of fragment $i$. $Q$ is the binding energy balance. Therefore, when $E_{avail} > 0$, the considered partition  is added to the list of all possible partitions which are energetically acceptable. It remains to sample the partitions according to their statistical weight. This is achieved by the usual phase space integrals~\cite{Amelin99}. The probability  $W(E_{avail},n)$ for each partition writes: 
\begin{equation}
 W(E_{avail},n) = \left(\frac{V}{\Omega}\right)^{n-1}\rho_n(E_{avail}),
\end{equation}
where $\rho_n(E_{avail})$ is the density in the final state. $V = 4\pi r_0^3\A/3$ is the volume of the decaying system and $r_0 = 1.3$~fm. $\Omega = (2\pi\hbar)^3$ is a normalization volume. $A$ is the mass number of the fragment to be deexcited. The density  $\rho_n(E_{avail})$ is the product of three terms:
\begin{equation}
\rho_n(E_{avail}) = M_n(E_{avail})S_nG_n.
\end{equation}
The first term is a phase space factor:
\begin{equation}
\label{equ:phasespace}
M_n(E_{avail}) = \int_{-\infty}^{+\infty}\cdots\int_{-\infty}^{+\infty}\delta\left( \sum_{i=1}^n\overrightarrow{\p_i} \right) \delta \left( E_{avail} - \sum_{i=1}^n \sqrt{\p_i^2+m_i^2}\right)\prod_{i=1}^n d^3\p_i,
\end{equation}
where $\overrightarrow{\p_i}$ is the momentum  of fragment $i$.

The second term is a spin factor taking into account the degeneracy of the considered state of fragment $i$:
\begin{equation}
S_n = \prod_{i=1}^n (2s_i+1).
\end{equation}
The last term is the usual combinatorial factor taking into account the multiplicity $n_i$ of each fragment $i$:
\begin{equation}
G_n = \prod_{j=1}^k \frac{1}{n_j!}.
\end{equation}
In the non relativistic case,  Equation \ref{equ:phasespace} has an analytical solution~\cite{Amelin99}. The probability writes:
\begin{equation}
W(E_{avail},n) = S_n G_n \left(\frac{V}{\Omega}\right)^{n-1} \left( \frac{\prod_{i=1}^n m_i}{\sum_{i=1}^n m_i} \right)^{3/2} \frac{(2\pi)^{3(n-1)/2}}{\Gamma(3(n-1)/2)} E_{avail}^{3n/2-5/2},
\end{equation}
where $m_i$ is the mass of fragment $i$.  

The choice of the partition is made by Monte-Carlo according to the statistical weight $W(E_{avail},n)$. The sampling of the momenta of each species is performed according to Equation~\ref{equ:phasespace} using the ROOT~\cite{ROOT} class {\it TGenPhaseSpace} based on the method of Raubold and Lynch~\cite{James68}.

For those fragments produced in the decay channel which are excited, a new iteration is produced until no more excited species are present in the final state.  Note that, since the system under study is rather light, the Coulomb interaction between all species in the final state is not taken into account. At the end of the de-excitation process,  a complete event is produced, boosted in the laboratory frame. The comparison with experimental data is then possible by taking into account the acceptance of the experimental set-up.  

\addtocontents{toc}{\vspace{0.4cm}}
\section{Comparisons with experimental data}
We now come to a comparison of our model with the experimental data obtained near the GANIL facility and presented in~\cite{Dudouet13b}. In the experiment, absolute cross-sections for light charged particles were measured over large angular and energy ranges. In a further experiment, additional measurements were also performed at $0$ degrees for some isotopes~\cite{Dudouet14b}.   
\subsection{Angular differential cross-sections}

\begin{figure*}[!ht]
\centering\includegraphics[width=\linewidth]{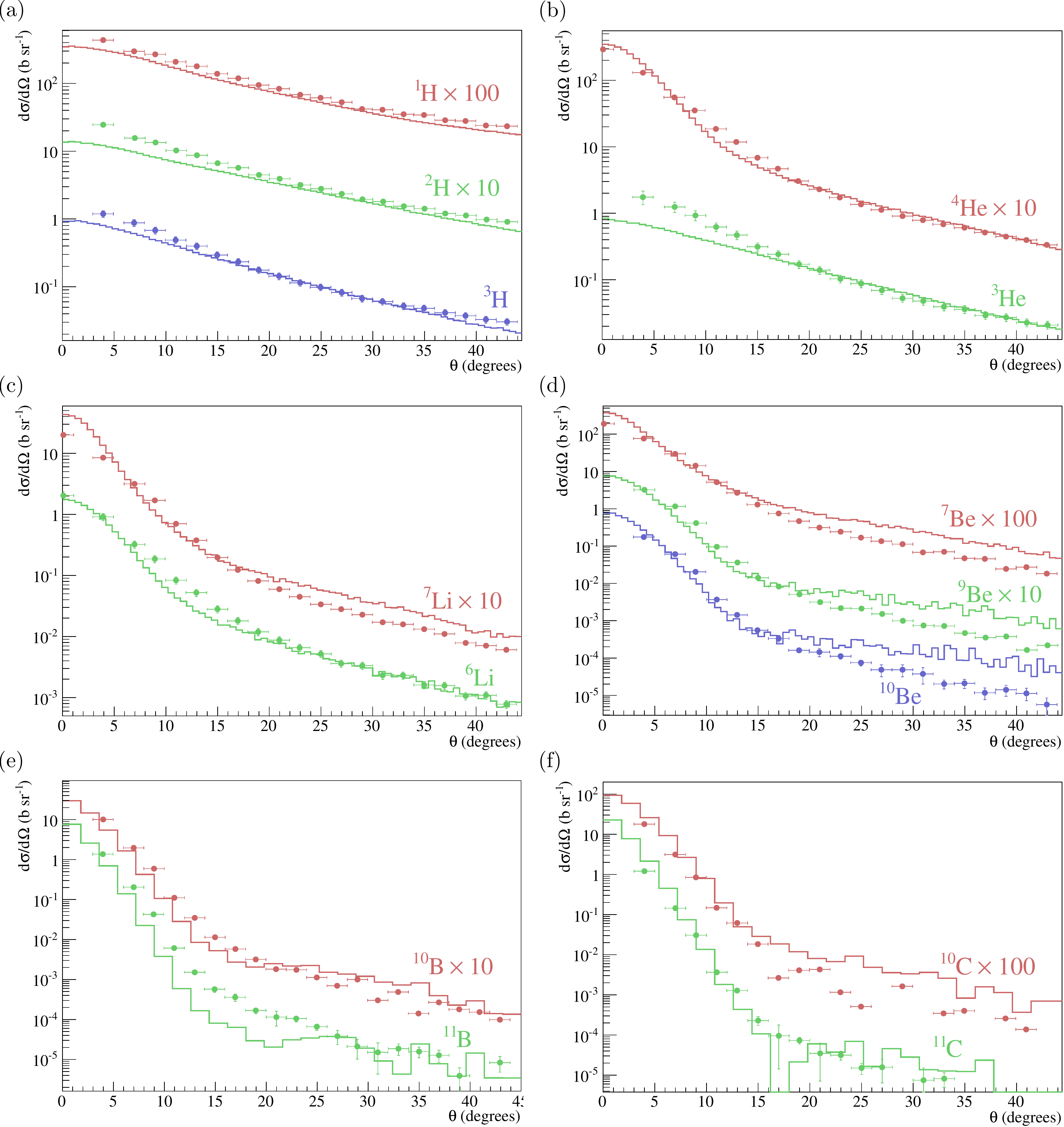}
\caption{(Color online) Angular differential cross-sections for various isotopes from $Z=1$ to $Z=6$. Points: experimental data from~\cite{Dudouet13b,Dudouet14b}. Histograms: results of the SLIIPIE model. For a sake of clarity, some values of the cross-sections have been multiplied by a factor indicated in each panel.}
\label{fig:MR_AngDist}
\end{figure*}

Figure~\ref{fig:MR_AngDist} shows the angular distributions as predicted by the SLIIPIE model (histograms). They are compared with the experimental data (points).  

The global features of the data are reproduced by the calculation although it is clear that the model particularly underestimates the production of deuterons, tritons and \noy{He}{3}{} at very forward angles. This part of the angular distribution is dominated by the decay of the projectile-like fragment. The underestimation is probably due to a lack of decay channels with emission of deuterons, tritons and \noy{He}{3}{}. These channels are associated with rather high excitation energies and the fact that excited stated in the continuum are not taken into account certainly leads to a lack of production. Since the measurements are limited to angles lower than $45$ degrees, the effect on the decay of the target-like fragment is not observed although it is apparent at least in the case of deuterons and tritons at the largest measured angles. Note that the model also slightly underpredicts the production of protons while it gives correct results for $\alpha$ particles. As far as fragments ($Z$ larger than 2) are concerned, results are quite good. In the model, these cross-sections results from the competition between the production of primary species and their subsequent secondary decay. The fact that a rather good agreement is obtained means that the production process is correctly taken into account. This process is essentially governed by geometrical effects since the size of the projectile-like and target-like fragments depends on the size of the participant zone which depends geometrically on the impact parameter. The overestimation of \noy{Li}{7}{} and \noy{Be}{7}{} yields at very forward angles may be due to the missing decay channels discussed above.        

\subsection{Double differential cross sections}

\begin{figure*}[!ht]
\centering\includegraphics[width=\linewidth]{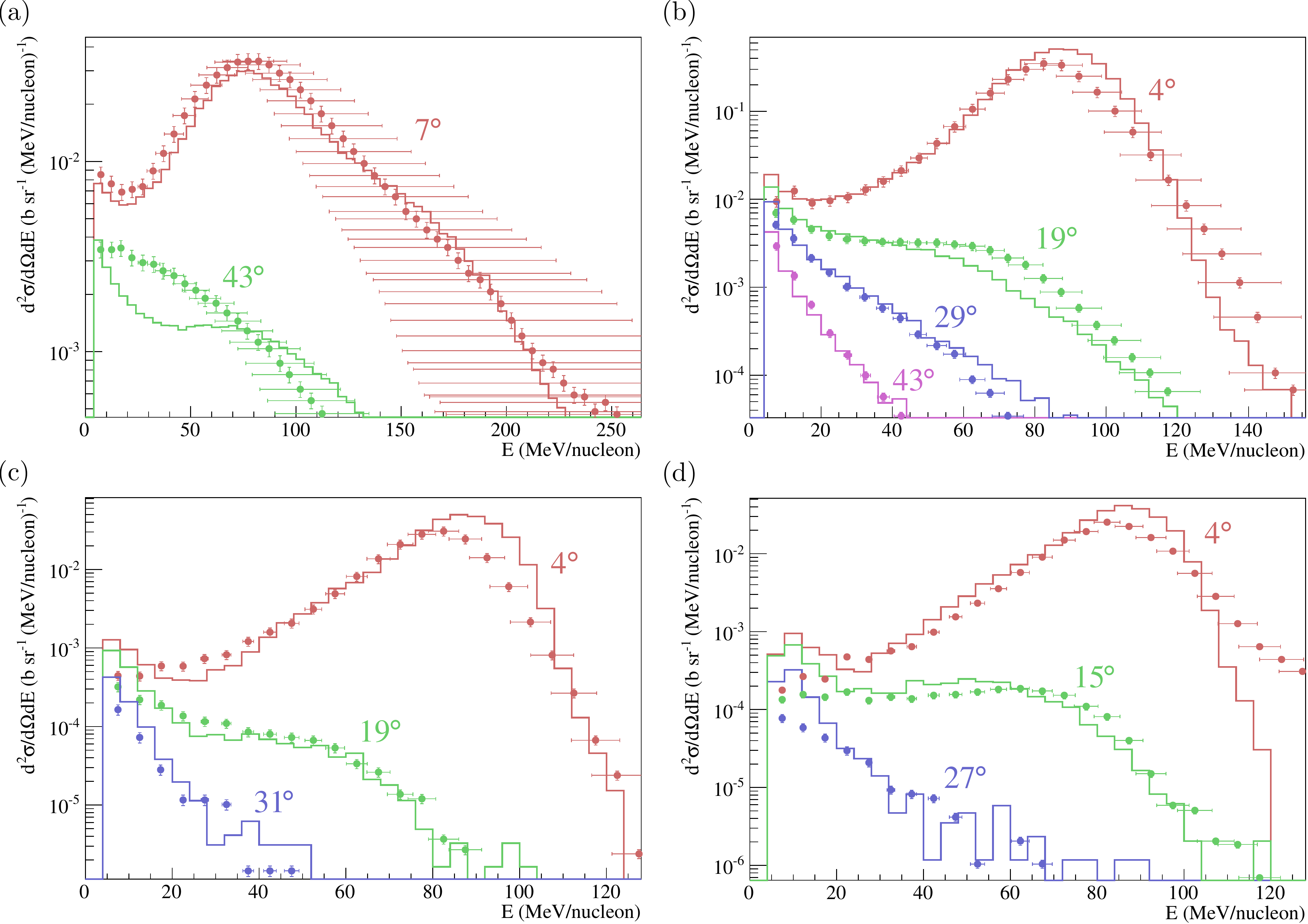}
\caption{(Color online) Double differential cross-sections for various isotopes: a) \noy{H}{1}{}, b) \noy{He}{4}{}, c) \noy{Li}{7}{}, d) \noy{Be}{7}{} detected at various angles indicated in the figure. The horizontal bars correspond to the uncertainty in the energy measurement (see~\cite{Dudouet13b} for more details).}
\label{fig:MR_EDist}
\end{figure*}

Figure~\ref{fig:MR_EDist} displays double differential cross sections for some isotopes and various angles.  Although discrepancies are observed in the proton spectra at large angles, the agreement for the other displayed  isotopes is satisfactory. At small angles, the kinetic energy spectra are dominated by the decay of the projectile-like fragment and the maxima of the distributions peaked at the beam energy are correctly reproduced. The model is also able to account for the energy distribution of clusters at large angles where the cross-section is dominated by processes occurring at mid-rapidity. Once again, this means that the assumptions regarding the physics in the overlapping zone of the two interacting nuclei is correct. The energy distributions slowly damp as the emission angle increases and the magnitude of the effect is reproduced both in shape and in magnitude. 

\subsection{Total production cross-sections}

\begin{figure*}[!ht]
\centering\includegraphics[width=\linewidth]{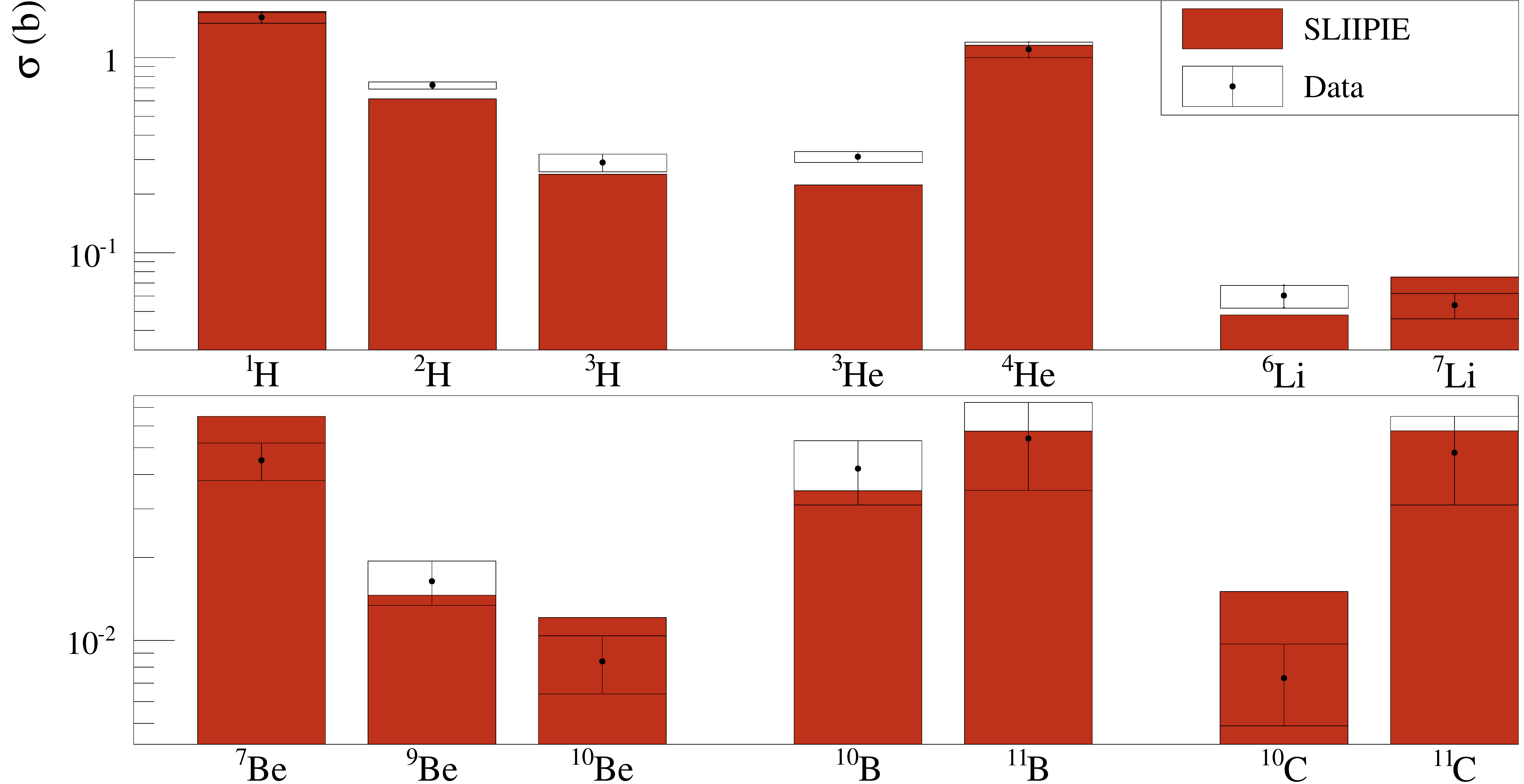}
\caption{(Color online) Total production cross-sections for various isotopes indicated in the figure: comparison between the prediction of SLIIPIE and the experimental data.}
\label{fig:sliipie}
\end{figure*}

\begin{figure*}[!ht]
\centering\includegraphics[width=\linewidth]{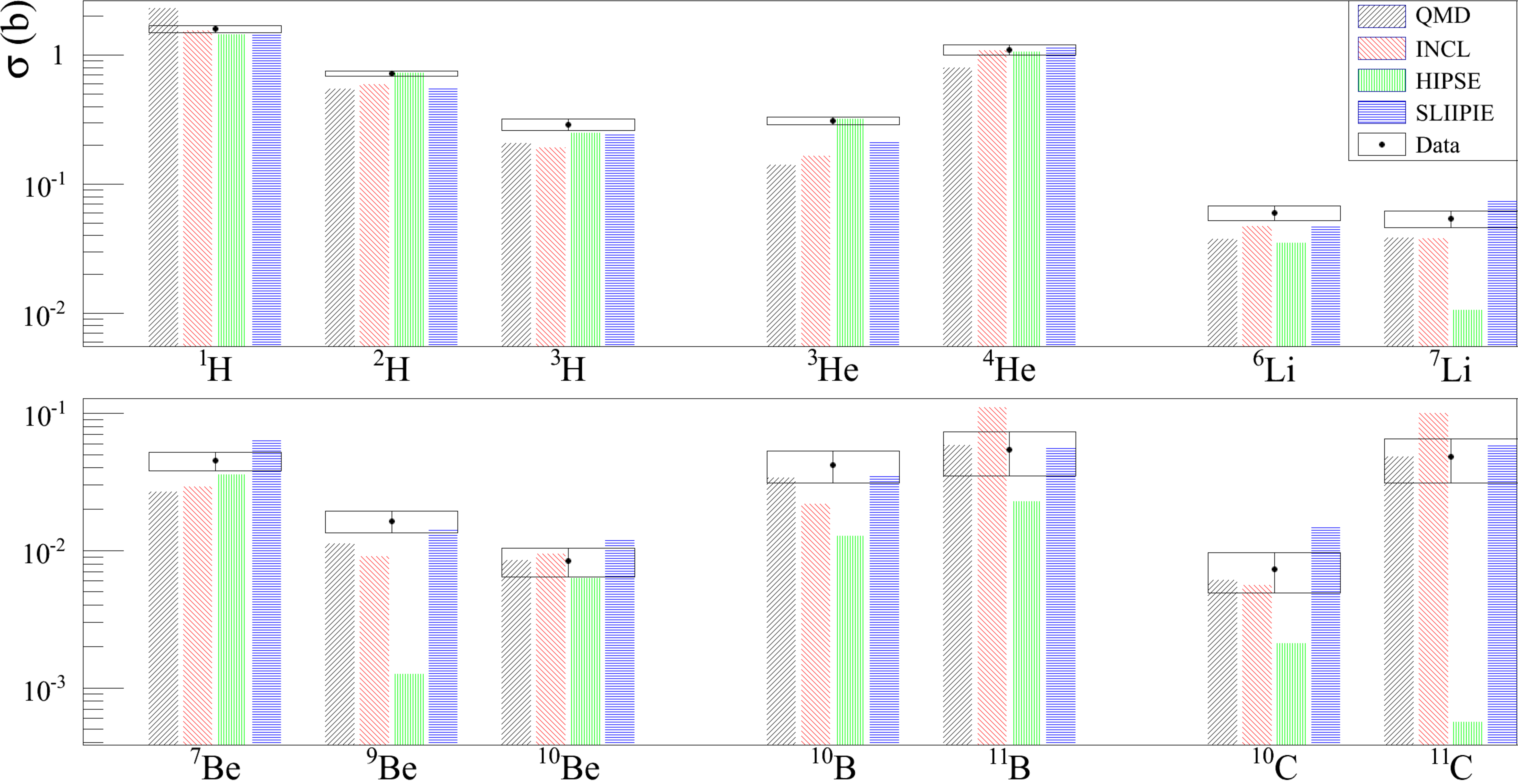}
\caption{(Color online) Comparison between the predictions of the models mentioned in the insert and the experimental results for several species.}
\label{fig:allmod}
\end{figure*}

\begin{figure}[!ht]
\centering\includegraphics[width=0.8\linewidth]{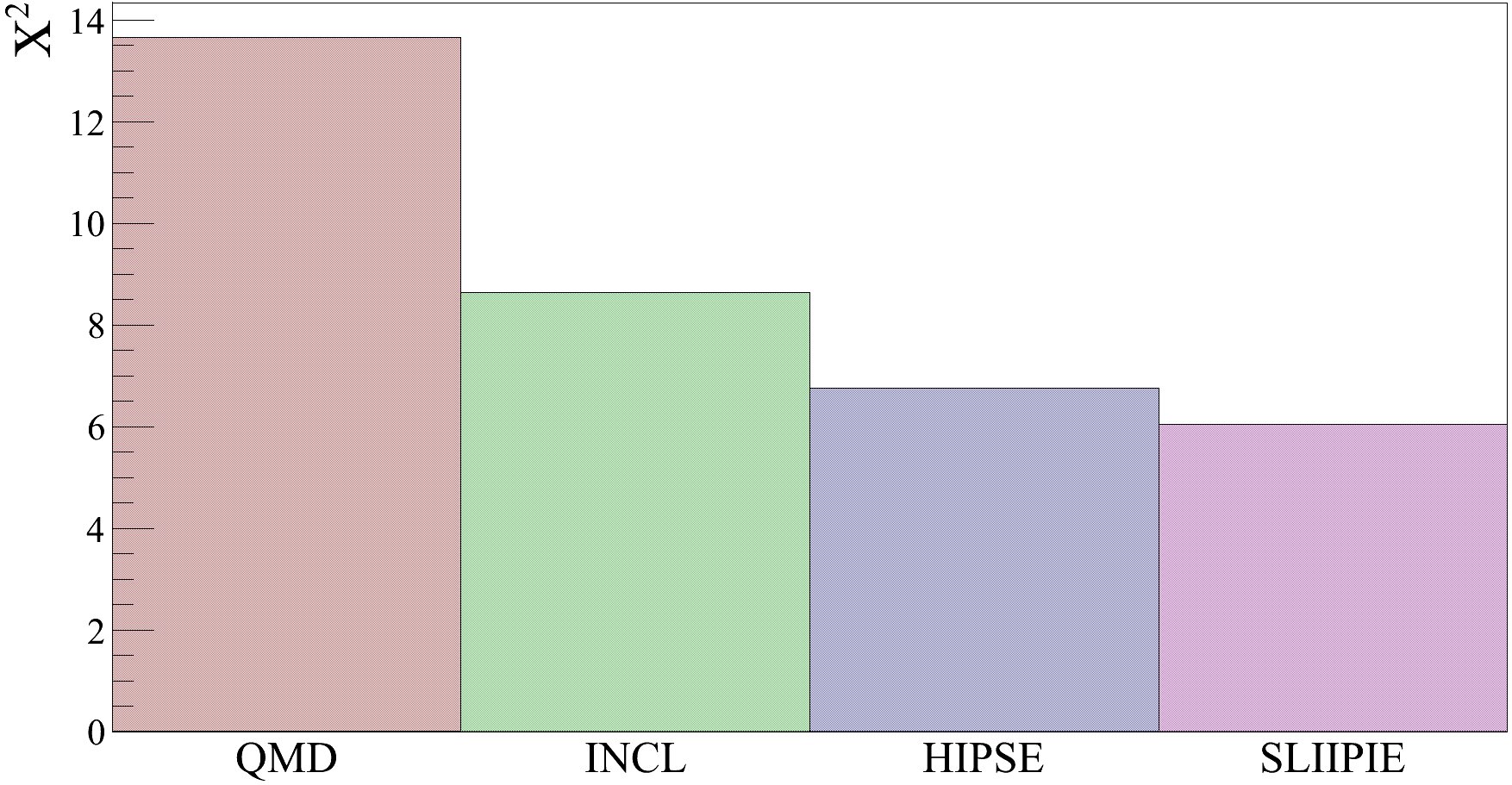}
\caption{(Color online) Comparison of the global Chi square for four different models.}
\label{fig:chi2}
\end{figure}

Total production cross-sections are displayed in Fig.~\ref{fig:sliipie}. A general agreement between experimental data and the results of the model is achieved for all considered species. A global comparison with other models available in the literature is shown in Fig.~\ref{fig:allmod}. The two first models are implemented in GEANT4: the Quantum Molecular Dynamics model~\cite{Koi10} (QMD) and the Intra-Nuclear Cascade of Liege model~\cite{Kaitaniemi11,Boudard13} (INCL). These models appeared to be the more predictive of the GEANT4 toolkit for this system in our previous benchmark~\cite{Dudouet14a}. The third one, HIPSE~\cite{HIPSE} (Heavy Ion Phase Space Exploration), is a phenomenological model which have been developed to simulate nuclear reactions around Fermi energies. 
Generally speaking, all models give the correct order of magnitude. In order to be more quantitative, a Chi-square including all species has been calculated for all considered model with respect to the experimental data. Although none of the models gives very accurate predictions, the SLIPIIE results are associated with the best Chi square (cf.~Fig.~\ref{fig:chi2}). 

\addtocontents{toc}{\vspace{0.4cm}}
\section{Conclusion}

In this work, we have presented a semi-microscopic model and compared the results with recent experimental data obtained in C-C reactions at 95 MeV/nucleon. From the rather successful agreement between experimental and calculated data, we may conclude that the main hypothesis of the model are valid. Our results point out a strong memory of the entrance channel characteristics of the reaction, namely the key role played by the participant-spectator geometrical assumption and the fact that the kinematics of the final products is governed to a large extent by the initial momentum distribution of the two partners of the reaction. This suggests a very fast clusterization process in the overlap region. Such a fast process is difficult to take into account in a fully dynamical description because it requires to take into account a very early coalescence mechanism. Usually, the dynamical models treat cluster production on longer time scales and this could be at the origin of the discrepancies between these models and the experimental data.  The ``sudden'' approximation used in our approach seems to be a key ingredient to reproduce the data, in particular in the mid-rapidity region. In future works, we plan to extend our model to other systems and other energies.

\bibliographystyle{unsrt}
\bibliography{SLIIPIE}

\end{document}